\documentclass[a4paper,11pt]{article}

\usepackage{graphicx}
\usepackage{dcolumn}
\usepackage{bm}
\usepackage{comment}
\usepackage{amssymb}
\usepackage{amsmath}

\begin{document}

\huge

\begin{center}
Equation of state of dense plasma mixtures: application to the Sun center
\end{center}

\vspace{1cm}

\large

\begin{center}
Jean-Christophe Pain\footnote{jean-christophe.pain@cea.fr} and G\'erard Dejonghe
\end{center}

\normalsize

\begin{center}
CEA, DAM, DIF, F-91297 Arpajon, France
\end{center}

\vspace{0.5cm}

\large

\begin{center}
Thomas Blenski
\end{center}

\normalsize

\begin{center}
CEA, DSM, IRAMIS, F-91191 Gif-sur-Yvette, France
\end{center}

\begin{abstract}
We present a self-consistent approach to the modeling of dense plasma mixtures in local thermodynamic equilibrium. In each electron configuration the nucleus is totally screened by electrons in a Wigner-Seitz sphere (ion-sphere model). Bound and free electrons are treated quantum-mechanically. The assumption that all species should have the same electronic environment leads to the equality of the electronic pressure for all ions of all elements having therefore different cell volumes. The variation of the average atomic radii of the different elements with respect to temperature is investigated, and the procedure is applied to the determination of pressure in the Sun center.
\end{abstract}

\section{Introduction}

Theoretical studies of electronic structure of disordered systems composed by ions and electrons are of great interest for understanding of dense-plasma radiative transfer and equation of state (EOS) \cite{BLENSKI07,PIRON11,PAIN07a}. In astrophysics, EOS of plasma mixtures is essential for stellar pulsation calculations, via pressure modes (requiring sound velocity given by the second derivative of the free energy with respect to matter density) or via gravity modes, highly sensitive to the temperature at the star center. Moreover, EOS has an impact on the photo-absorption cross-sections, which determine temperature distribution inside the star. Such EOS studies concern atoms in extreme conditions of density and temperature, which implies existence of multicharged ions close to local thermodynamic equilibrium (LTE). An ion configuration is defined by sub-shells (orbitals) occupied by an integer number of bound electrons. For example $c=(1s)^2(2s)^2(2p)^5(3s)^1(3p)^3(3d)^2$ is a configuration consisting of six sub-shells containing respectively 2, 2, 5, 1, 3 and 2 electron(s). In our model, the nucleus is totally screened by bound and free electrons in a sphere (ion-sphere model) which is usually named, refering to solid-state physics, a Wigner-Seitz (WS) sphere. Throughout the paper, the term ``element'' means ``element of the periodic table'' (chemical element): H, He, Li, Be, B, C, $\cdots$. 

In the usual so-called ideal-gas mixing rule, the electronic structure of each element is calculated separately and a mixture is obtained by simply including all the individual elements at fixed temperature with their abundances. There are also other ways to treat mixtures, for instance in the framework of orbital-free molecular dynamics \cite{LAMBERT08}.

The present model relies on the idea that at the LTE conditions, all ionic species shall be considered at the same plasma electronic environment (free electron "bath"). In section \ref{sec1}, we show that, provided that ions of different charges have different volumes, the electronic pressure, calculated together with ionic volumes in a self-consistent way, must be equal for each ion of each element. Comparisons with chemical-picture \cite{POTEKHIN10} and average-atom mixture models \cite{YUAN02} in the case of a CO$_2$ plasma are presented in section \ref{sec2}. In the same section, values of pressure at the Sun center are compared with the results of OPAL physical-picture model \cite{ROGERS86,ROGERS96,ROGERS02} and Quantum Langevin Molecular Dynamics (QLMD) simulations \cite{DAI10}. 

\begin{figure}
\vspace{1cm}
\begin{center}
\includegraphics[width=10cm]{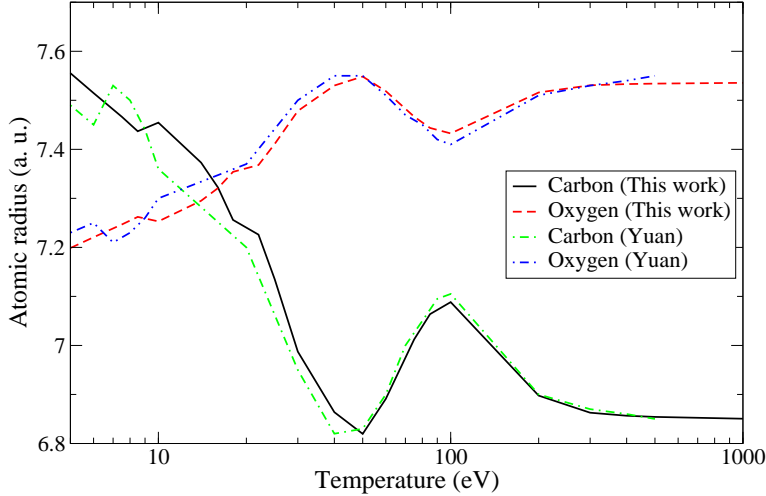}
\end{center}
\caption{Variation of atomic radii of carbon and oxygen with respect to the temperature and comparison with the results of Yuan \cite{YUAN02}. The non-monotonicity illustrates the competition between electronic structures of the two atoms.}\label{fig1}
\end{figure}

%=======================================================
%			I. ATOMIC STRUCTURE AND PRESSURE
%=======================================================

\section{\label{sec1} Self-consistent-field calculation of atomic structure and equality of electronic boundary pressure}

Atoms in a plasma can be idealized by an average atom confined in a Wigner-Seitz (WS) sphere, which radius $r_{ws}$ is related to matter density. The electronic pressure $P_e$ \cite{PAIN06a,PAIN06b} consists of three contributions, $P_e=P_{b}+P_{f}+P_{xc}$, where the bound- and free- electron pressures, respectively $P_b$ and $P_f$, are evaluated using the stress-tensor formula

\begin{equation}\label{presf2}
P_{b,f}=\frac{1}{8\pi r_{ws}^2}\sum\!\!\!\!\!\!\!\!\int\frac{f(\epsilon_s,\mu)}{\left(1+\frac{\epsilon_s}{2E_0}\right)}\left[\left(\frac{dy_s}{dr}\Big|_{r_{ws}}\right)^2+ \left(2\epsilon_s\left(1+\frac{\epsilon_s}{2E_0}\right)-\frac{1+\ell+\ell^2}{r_{ws}^2}\right)y_s^2(r_{ws})\right],
\end{equation}

$y_s$ representing the radial part of the wavefunction $\psi_s$ (corresponding to energy $\epsilon_s$) multiplied by $r$, $E_0$ the rest-mass energy of the electron and $\ell$ the orbital angular momentum. $f$ denotes the Fermi-Dirac distribution:

\begin{equation}
f(\epsilon_s,\mu)=\frac{2(2\ell+1)}{1+e^{\underline{}\beta(\epsilon_s-\mu)}},
\end{equation}

$\mu$ being the chemical potential. For bound states, $s\equiv n,\ell$ and $\sum\!\!\!\!\!\!\int\equiv\sum_{n=1}^{\infty}\sum_{\ell=0}^{n-1}$, and for free states, $s\equiv\epsilon,\ell$ and $\sum\!\!\!\!\!\!\int\equiv\int_0^{\infty}d\epsilon_s\sum_{\ell=0}^{\infty}$. The choice of the boundary conditions plays a major role in the pressure value. This comes from the fact that the energy of an orbital depends on the value of the corresponding wavefunction at the boundary of the WS sphere. In our model, the bound wavefunction behaves like a decreasing exponential at the boundary. $P_{xc}$ is the exchange-correlation pressure evaluated in the local density approximation using the formulas of Ichimaru \emph{et al.} \cite{ICHIMARU87}. The adiabatic approximation is used to separate the thermodynamic functions into electronic and ionic components. The total pressure can be written $P=P_i+P_e=P_i+P_b+P_f+P_{xc}$, where $P_i$ is the ionic pressure, calculated according to \cite{NIKIFOROV87} (ideal gas with One-Component-Plasma corrections). The self-consistent calculation of a configuration provides the potential, the electron populations of the orbitals and the free-electron chemical potential $\mu$, resulting from the neutrality of the plasma.

In the plasma, all ions should have the same electronic environment. All configurations of the same charge are grouped together and constitute an ion. Normalization of configuration probabilities ($\sum_cW_c=1$ associated to Lagrange multiplier $B$) and preservation of total volume ($\sum_cW_cV_c=V$ associated to Lagrange multiplier $\bar{P}$) must be ensured. The constrainted grand potential is:

\begin{equation}
\tilde{F}=\sum_cW_cF_c+B\left[\sum_cW_c-1\right]+\bar{P}\left[\sum_cW_cV_c-V\right]+k_BT\sum_cW_c\ln W_c,
\end{equation}

and the variation with respect to volume $V_c$ and probability $W_c$ give respectively 

\begin{equation}\label{press1}
P_c=-\frac{\partial F}{\partial V_c}\Bigl|_{T}=\bar{P}=P\;\;\;\; \text{and}\;\;\;\; W_c\propto\exp\left[-\frac{\tilde{F}_c+PV_c}{k_BT}\right].
\end{equation}

Equalities of Eq. (\ref{press1}) mean \cite{PAIN02b,PAIN03} that all configurations have the same pressure and that it is necessary to consider enthalpy instead of the free energy. In practical, all pressures are calculated in a self-consistent way and equalized by a multi-dimensional Newton-Raphson method. The average value of thermodynamic quantity $X(e,i)$ for ion $i(e)$ of element $e$ and its probability $\tilde{W}_{i(e)}$ are given by:

\begin{equation}
X(e,i)=\frac{\sum_{c\in i(e)}W_cX(c)}{\sum_{c\in i(e)}W_c},\;\;\;\; \text{and}\;\;\;\; \tilde{W}_{i(e)}=\frac{\sum_{c\in i(e)}W_c}{\sum_cW_c},
\end{equation}

where $W_c$ and $X(c)$ represent respectively the probability and the value of quantity $X$ for configuration $c$. The average value $\langle X\rangle$ of a quantity $X$ thus reads:

\begin{equation}
\langle X\rangle=\sum_{e=1}^{E}\sum_{i(e)=1}^{N_e}\tilde{W}_{i(e)}X(e,i)=\frac{\sum_{e=1}^{E}\sum_{i(e)=1}^{N_e}\sum_{c\in i(e)}W_{c}X(c)}{\sum_cW_c},
\end{equation} 

where $E$ is the number of elements in the plasma and $N_e$ the number of charge states of element $e$. In the present work, pressure equality means the equality of the stress-tensor pressure at the boundary of all ionic spheres. Similarly to the single-element case \cite{PAIN02b,PAIN03}, it is convenient to write the system of equations in the following form: $P_{1(1)}=P_{2(1)}$, $P_{2(1)}=P_{3(1)}$, $\cdots$, $P_{N_1(1)}=P_{1(2)}$, $P_{1(2)}=P_{2(2)}$, $P_{2(2)}=P_{3(2)}$, $\cdots$, $P_{N_2(2)}=P_{1(3)}$, $\cdots$, $P_{1(E)}=P_{2(E)}$, $\cdots$, $P_{N_E-1(E)}=P_{N_E(E)}$, where $P_{i(e)}$ represents pressure at the boundary of ion $i$ of element $e$. The partial density of this ion is $\rho_{i(e)}$ and the index $i(e)$ varies from $1$ to $N_e$. The closure relation of the system is the matter conservation law

\begin{equation}
\sum_{e=1}^{E}\sum_{i(e)=1}^{N_e}\frac{\tilde{W}_{i(e)}x_eM_e}{\rho_{i(e)}}=\frac{\sum_{e=1}^{E}x_eM_e}{\rho_t},
\end{equation}

where $\rho_{t}$ is the matter density of the plasma and $M_e$ and $x_e$ represent respectively the atomic mass and the mass fraction of element $e$. The Jacobian matrix can be inverted analytically since only diagonal-, first super-diagonal- and last-line coefficients are non zero. 

\vspace{0.5cm}

\begin{table}
\begin{center}
\begin{tabular}{|c|c|c|} \hline \hline
 & Total pressure & Total pressure \\ 
 T & Potekhin and Chabrier & This work \\
 (eV) & (Mbar) & (Mbar) \\ \hline \hline
50 & 2.22 & 1.86 \\\hline
100 & 5.06 & 4.46 \\\hline
500 & 27.16 & 27.17 \\\hline
\end{tabular}
\end{center}
\caption{Total pressure for a CO$_2$ plasma at $\rho_t$=0.1 g/cm$^3$. Comparison with values obtained from the chemical-picture model of Potekhin and Chabrier \cite{POTEKHIN10}.}\label{tab1}
\end{table}

\begin{table}
\begin{center}
\begin{tabular}{|c|c|c||c|c|c|c|c|} \hline \hline
Element & Z$^*$ & $\rho$ & $P_e$ (Mbar) & Element & Z$^*$ & $\rho$ & $P_e$ (Mbar) \\ \hline \hline
 H & 1 & 105.81 & 2.10 \; 10$^5$ & P & 13.84 & 248.12 & 8.50 \; 10$^4$ \\\hline
 He & 2 & 211.74 & 1.03 \; 10$^5$ & S & 14.62 & 243.73 & 8.67 \; 10$^4$ \\\hline
 C & 5.69 & 228.02 & 9.64 \; 10$^4$ & Cl & 14.63 & 269.70 & 8.29 \; 10$^4$ \\\hline
 N & 6.62 & 229.25 & 9.63 \; 10$^4$ & Ar & 15.29 & 291.31 & 7.78 \; 10$^4$\\ \hline
 O & 7.57 & 229.68 & 9.61 \; 10$^4$ & Ca & 16.86 & 265.98 & 8.47 \; 10$^4$\\ \hline
 Ne & 9.47 & 233.11 & 9.47 \; 10$^4$ & Ti & 18.52 & 290.34 & 7.77 \; 10$^4$\\ \hline
 Na & 10.41 & 242.45 & 9.11 \; 10$^4$ & Cr & 20.23 & 289.58 & 7.77 \; 10$^4$\\\hline
 Mg & 11.33 & 236.13 & 9.35 \; 10$^4$ & Mn & 21.09 & 293.99 & 7.62 \; 10$^4$\\\hline
 Al & 12.23 & 243.33 & 9.01 \; 10$^4$ & Fe & 21.95 & 287.63 & 7.76 \; 10$^4$\\\hline
 Si & 13.13 & 236.62 & 9.20 \; 10$^4$ & Ni & 23.21 & 281.42 & 7.73 \; 10$^4$ \\ \hline
\end{tabular}
\end{center}
\caption{Partial ionizations $Z^*$ and densities $\rho$ (in g/cm$^3$) obtained from our mixture model with Grevesse composition \cite{GREVESSE84}. $P_e$ represents the pressure of each element before applying the mixture procedure.}\label{tab2}
\end{table}

\begin{table}
\begin{center}
\begin{tabular}{|c|c|c|c|} \hline \hline
$\rho_t$ (g/cm$^3$) & $T$ (eV) & EOS model & $P$ (Mbar) \\ \hline \hline
141.25 & 989.45 & OPAL & 1.6162 \; 10$^5$ \\ \hline
       &        & QLMD & 1.6090 \; 10$^5$ \\ \hline
       &        & This work & 1.6217 \; 10$^5$ \\ \hline
141.25 & 1189.58  & OPAL & 1.9337 \; 10$^5$ \\ \hline
       &          & QLMD & 1.9122 \; 10$^5$ \\ \hline
       &          & This work & 1.9403 \; 10$^5$ \\ \hline
152.70 & 1352.64  & QLMD$^a$ & 2.3525 \; 10$^5$ \\ \hline
       &          & This work$^a$ & 2.3584 \; 10$^5$ \\ \hline
       &          & QLMD$^b$ & 2.2575 \; 10$^5$ \\ \hline
       &          & This work$^b$ & 2.2629 \; 10$^5$ \\ \hline
       &          & QLMD$^c$ & 2.2328 \; 10$^5$ \\ \hline
       &          & This work$^c$ & 2.2532 \; 10$^5$ \\ \hline                      
\end{tabular}
\end{center}
\caption{Comparison of central pressures of the Sun at typical temperatures ($T$) and densities ($\rho$) between our approach, the OPAL equation of state \cite{ROGERS86,ROGERS96,ROGERS02} and the Quantum Langevin Molecular Dynamics \cite{DAI10}. $(^a)$ corresponds to the mass fractions of H, He and C: X=0.3387, Y=0.6613 and Z=0, $(^b)$ corresponds to the mass fractions of H, He and C: X=0.3125, Y=0.6406 and Z=0.0469 and $(^c)$ corresponds to the mass fractions of H, He and O: X=0.3077, Y=0.6308 and Z=0.0615.}\label{tab3}
\end{table}

\begin{figure}
\vspace{1cm}
\begin{center}
\includegraphics[width=12cm]{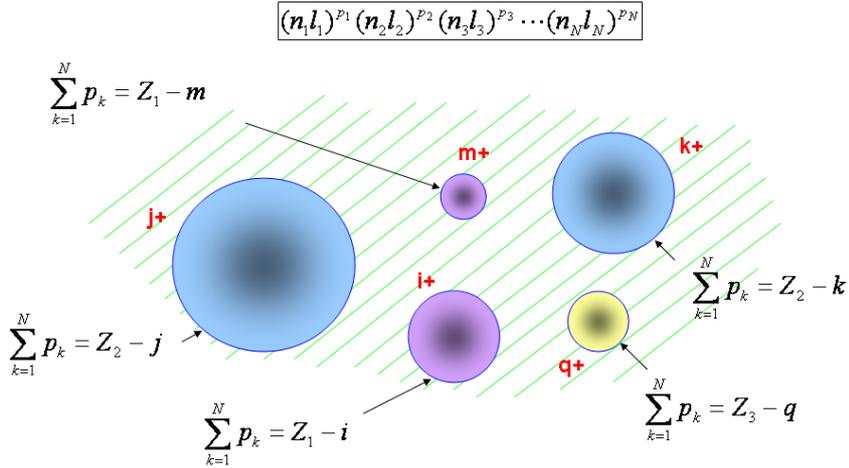}
\end{center}
\caption{Schematic picture of the statistical mixture model.}\label{fig2}
\end{figure}

\section{\label{sec2} Comparisons with other approaches and application to the Sun center}

Figure \ref{fig1} shows, in the case of a CO$_2$ plasma at a density of 0.1 g/cm$^3$ \cite{YUAN02}, that the atomic radii of carbon and oxygen exhibit a non-monotonic variation with the temperature. This is the signature of the shell structure, absent from Thomas-Fermi model \cite{FEYNMAN49}. The competition between the ionization of the two atoms can be seen very clearly. One can find that there are two main structures around 10 and 50 eV, respectively. The first increase of the carbon radius around 10 eV is due to the ionization of the L shell ($2s$ and $2p$ orbitals). For oxygen, this occurs later due to the fact that the binding energies are higher than for carbon. The second increase of radii, around 50 eV, is due do the ionization of the K shell ($1s$ orbital) which is ionized more rapidly for carbon than for oxygen. The radii obtained from the average-atom mixture model of Yuan \cite{YUAN02} differ from ours mainly at low temperatures. Total pressures are compared with the values of Potekhin and Chabrier \cite{POTEKHIN10} for three different temperatures in table \ref{tab1}.

The solar neutrino flux predicted by a solar model depends on the Rosseland opacity of the stellar material in the Sun center \cite{TURCK93}. Opacity influences the details of nuclear reactions and therefore the solar luminosity. Heliosismology provides a large array of accurately measured (0.01 \%) p-mode oscillation frequencies, related to the equation of state of the solar plasma. Temperature is taken to be $T_e$=1.3621 keV and density $\rho_t$=157.02 g/cm$^3$ and the abundances of different elements are taken from the well-known Grevesse mixture \cite{GREVESSE84}. The electron pressure predicted by our model is 1.4 Mbar and the total pressure 2.43 Mbar (if ions are treated as a pure ideal gas) and 2.46 Mbar (if OCP corrections are included). The resulting partial ionizations and densities are displayed in table \ref{tab2}, and comparisons with the OPAL values \cite{ROGERS86,ROGERS96,ROGERS02} and with the QLMD results of Dai \emph{et al.} in table \ref{tab3}. The OPAL equation of state belongs to the ``physical-picture'' models, and relies on an activity expansion. The bound states are described through the Planck-Larkin partition function, and the systematic expansion from one-body to many-body terms allows one to take into account the continuum contribution (scattering states) \cite{DAPPEN10}. Solar mixture can be a good test for our models: for instance very small quantities of Fe and Ni provide a large contribution to the opacity, which influences the details of nuclear reactions (and therefore the solar luminosity) as well as the predicted neutrino flux \cite{TURCK93}. 

\section{Conclusion}

A new approach for the thermodynamics of plasma mixtures has been presented. All electrons are treated quantum-mechanically and the shell structure has an important impact on the partial densities of different elements. The electronic pressure is imposed to be equal at the boundary of each ion sphere. Thus, the partial density of each ions charge state is determined in a self-consistent way. Future work includes further comparisons with other approaches, such as chemical-picture models \cite{POTEKHIN10,LOBODA09} and improvement of thermodynamic consistency (see Refs. \cite{BLENSKI07,PIRON11}). We plan also to study other thermodynamic quantities, such as internal energy, sound speed or gravitational acceleration, and to apply the present model to the calculation of electrical resistivity \cite{PAIN10} of plasma mixtures.

\end{document}